\documentclass[
 a4paper,twocolumn,11pt,accepted=2022-10-03
]{quantumarticle}
\pdfoutput=1
\usepackage[utf8]{inputenc}
\usepackage{amsmath}
\usepackage{amsfonts}
\usepackage{comment}
\usepackage{amssymb}
\usepackage{graphicx}
\usepackage{physics}
\usepackage{braket}
\usepackage{comment}
\usepackage{xcolor}
\usepackage{float}
\usepackage{tikz}
\usepackage{hyperref}
\usetikzlibrary{calc,positioning,decorations.pathreplacing,calligraphy,matrix,graphs,fit,shapes,arrows.meta}
\tikzset{>=latex}
\usepackage[left=2cm,right=2cm,top=2cm,bottom=2cm]{geometry}
\usepackage[left=2cm,right=2cm,top=2cm,bottom=2cm]{geometry}
\usepackage[backend=bibtex,style=phys,biblabel=brackets,
chaptertitle=false,pageranges=false,eprint=true,doi=false]{biblatex}

\DeclareFieldFormat
  [article,inbook,incollection,inproceedings,patent,thesis,unpublished,book,misc]
  {title}{#1\isdot}

\DeclareMathOperator{\Tra}{Tr}
\usepackage[export]{adjustbox}

\begin{document}
\newcommand{\sket}[1]{\ket{#1}\rangle}
\newcommand{\sbra}[1]{\langle\bra{#1}}
\newcommand{\sbraket}[1]{\langle\braket{#1}\rangle}
\newcommand{\exv}[1]{\left\langle #1 \right\rangle}
\newcommand{\timo}{\overleftarrow{T}}

\newcommand{\ig}[1]{\includegraphics[width=\linewidth]{#1}}
\title{Using the Environment to Understand non-Markovian Open Quantum Systems}
\author{Dominic Gribben}
\affiliation{SUPA, School of Physics and Astronomy, University of St Andrews, St Andrews KY16 9SS, United Kingdom}
\author{Aidan Strathearn}
 \affiliation{School of Mathematics and Physics, The University of Queensland, St Lucia,
Queensland 4072, Australia}
\author{Gerald E. Fux}
\affiliation{SUPA, School of Physics and Astronomy, University of St Andrews, St Andrews KY16 9SS, United Kingdom}
\author{Peter Kirton}
\affiliation{Department of Physics and SUPA, University of Strathclyde, Glasgow G4 0NG, United Kingdom}
\author{Brendon W. Lovett}
\affiliation{SUPA, School of Physics and Astronomy, University of St Andrews, St Andrews KY16 9SS, United Kingdom}

\begin{abstract}
Tracing out the environmental degrees of freedom is a necessary procedure when simulating open quantum systems. While being an essential step in deriving a tractable master equation it represents a loss of information. In situations where there is strong interplay between the system and environmental degrees of freedom this loss makes understanding the dynamics challenging. These dynamics, when viewed in isolation, have no time-local description: they are non-Markovian and memory effects induce complex features that are difficult to interpret.
To address this problem, we here show how to use system correlations, calculated by any method, to infer any correlation function of a Gaussian environment, so long as the coupling between system and environment is linear. This not only allows reconstruction of the full dynamics of both system and environment, but also opens avenues into studying the effect of a system on its environment.
In order to obtain accurate bath dynamics, we exploit a numerically exact approach to simulating the system dynamics, which is based on the construction and contraction of a tensor network that represents the process tensor of this open quantum system.
Using this we are able to find any system correlation function exactly.  To demonstrate the applicability of our method we show how heat moves between different modes of a bosonic bath when coupled to a two-level system that is subject to an off-resonant drive.
\end{abstract}

\maketitle

\section{Introduction \label{sec:intro}}
When modelling a realistic quantum system the effect of its environment must be captured in some form~\cite{breuer2002theory}. The environment consists of many degrees of freedom and is often approximated as an infinite collection of harmonic oscillators. Thus, an explicit description of the environment is usually impossible. Most methods for describing open quantum systems rely on tracing out the environment to obtain an effective description of the system involving only a tractable number of degrees of freedom. When the system-environment coupling is strong, and/or when the spectral density of the environment is structured, the full dynamics involves a significant interplay between the system and environment degrees of freedom~\cite{chin2013role, thorwart2009enhanced, mohseni2014energy, rey2013exploiting, maier2019environment, segal2016vibrational, groblacher2015observation, potovcnik2018studying, Xiong2018PhotonicCW}. When the environment is traced out, the dynamics of the system cannot be described by a time-local master equation and in this sense are  non-Markovian~\cite{de2017dynamics,breuer2016colloquium}, whose simulation generally requires sophisticated numerical methods. In addition, such system dynamics are often complex to interpret. 


Analyzing the behaviour of the environment is useful for providing insights into processes where the  system-environment interaction is non-trivial or for any application where the goal is to manipulate the environment via the system. An important example of this class of problem occurs in modelling quantum thermal machines~\cite{mitchison2019quantum,binder2018thermodynamics} where distinct thermodynamic effects beyond weak coupling can be observed~\cite{brenes2020tensor,strasberg2016nonequilibrium}. Access to the state of the environment could also be used to explicitly track the formation of bound states such as polarons~\cite{mahan2013many,silbey1984variational,harris1985variational,devreese2009frohlich} or polaritons~\cite{mahan2013many,kavokin2003cavity}. There is also potential for simulating phase transitions in the environment, such as to lasing or superradiant states~\cite{kirton2018superradiant} in multimode cavity QED~\cite{kollar2017}. As well as these, the technique we introduce here allows for calculation of out-of-time-order correlations of the environment; these are known to characterise the degree of information scrambling in many-body quantum systems~\cite{garttner2017measuring,li2017measuring,niknam2020sensitivity}.

Some existing techniques allow varying levels of insight into the behaviour of the environment. These generally rely on a direct calculation of the dynamics of the environment itself. For example, the reaction coordinate mapping explicitly calculates the time-dependence of a collective coordinate that gives a sense of the behaviour of the environment~\cite{iles2014environmental,iles2016energy}. It has also been shown that the auxiliary density operators calculated in the hierarchy equations of motion method can be used to calculate the same collective coordinate dynamics~\cite{lambert2019modelling} as well as higher order moments of the total bath coupling operator~\cite{zhu2012explicit,song2017hierarchical,schinabeck2018hierarchical}. Counting field techniques can also be used to calculate moments of the total bath coupling operator~\cite{popovic2020non,esposito2009nonequilibrium,kilgour2019path}.  Techniques such as TEDOPA~\cite{prior10} map the star topology of an environment to a chain of modes with nearest neighbour couplings. Dynamics within this chain can also be used to characterise the behaviour of the environment~\cite{tamascelli2020excitation} and these dynamics can be passed through an inverse chain-mapping to extract the full environmental dynamics~\cite{schroder2016simulating,gonzalez2017uncovering}.

In contrast here we give a prescription that allows calculation of any multi-time bath correlation function using only system moments of equal order: no auxiliary modes or additional degrees of freedom are required. Our method can therefore be used in conjunction with any technique that allows for the accurate calculation of system correlation functions, in order to yield bath dynamics and correlations. This is even possible using those methods where the entire environment is traced out.

Accurately calculating the required correlation functions for a system undergoing non-Markovian evolution is numerically challenging for all but the simplest cases. It is necessary to use a technique which is capable of recording the history of the system dynamics while allowing for the application of arbitrary sequences of system operators. Here we employ a version of the time-evolving matrix product operator (TEMPO) algorithm~\cite{strathearn2017efficient}. This has been shown to efficiently capture the required time non-local influences by representing the discretised trajectories  of the system as a matrix product state (MPS). This is very similar to the use of MPS in simulating 1D quantum many-body problems which can be compressed by truncating the sizes of the matrices, keeping only the most relevant contributions. An imaginary time version of TEMPO can be used to determine the thermal state of a system with a strongly-coupled environment~\cite{chiu2021numerical}.
However, even with the TEMPO algorithm, calculating all of the necessary correlation functions would require many separate numerical simulations. This difficulty can be overcome by building on the approach developed in~\cite{Jorgensen2019exploiting} where a modification of the network contracted in the TEMPO algorithm allowed efficient generation of an object called the process tensor. From a single process tensor it is possible to access not just the system dynamics but also all system correlation functions at arbitrary times~\cite{fux2021efficient}. This significantly reduces the computational overhead of calculating the behaviour of the environment.  

In a previous work~\cite{gribben2020exact} we showed how the displacement of a bosonic bath that is linearly coupled to a spin system can be calculated from a simple transformation of the system dynamics. We used this to gain insight into the origin of some of the non-Markovian features in the dynamics. The approach used there is limited to the calculation of first moments. In Sec.~\ref{sec:BD} of this paper we reformulate the problem in terms of a moment generating functional, to show how we can calculate all higher order bath correlation functions from system correlations alone. 
An overview of the process tensor formulation~\cite{Pollock2018, fux2021efficient} of open quantum systems is given in Sec.~\ref{sec:proctens} where we demonstrate how it can be used to efficiently calculate the system moments necessary for calculation of higher order bath correlation functions. Finally, in Sec.~\ref{sec:SBM} we use this approach to study how the biased spin-boson model~\cite{leggett1987dynamics} can be tuned to redistribute energy in its environment.

\section{Bath Dynamics \label{sec:BD}}

Here we provide a method to calculate any correlation function of bath operators in terms of those of system operators, for a system linearly coupled to a Gaussian environment. We do this by expressing the relevant expectation values of environment operators as derivatives of a generating functional.
By adding source terms to the propagator that remain easy to integrate out for the Gaussian bath, the derivatives required to calculate the relevant moments can then be carried out explicitly.

The system-environment Hamiltonian we consider has the general form
\begin{align}\label{eq:basicham}
H &= H_S + H_I + H_B \nonumber
\\ &= H_S+s \sum_q g_q (a_q + a_q^\dagger) + \sum_q \omega_q a_q^\dagger a_q,
\end{align}
where $a_q$ destroys an excitation in bosonic environment mode $q$ with frequency $\omega_q$.
Here, $s$ is a general operator acting exclusively on the system Hilbert space and $g_q$ is the coupling strength of mode $q$ to the system. Formally, the full dynamics, in the interaction picture, can be written as~\cite{mukamel1995principles}:
\begin{equation}\label{eq:eom}
\rho(t) = \timo  \exp\left[\int_0^t  \mathcal{L}_I(t') \, dt' \right] \rho(0).
\end{equation}
The full density operator of the system and environment, is $\rho(t)$ and $\mathcal{L}_I\rho=-i[H_I(t), \rho]$ is the Liouvillian superoperator corresponding to unitary dynamics under the interaction Hamiltonian and $\timo$ indicates that superoperators are to be time-ordered from right to left. Details of the superoperator formalism are given in App.~\ref{app:sup}.

To proceed we assume that the initial state is factorizable and can be written as \mbox{$\rho (0) = \rho_S (0)\otimes \rho_B (0)$}. We can then trace out the bath and arrive at an expression for the evolution of the system:
\begin{equation}\label{eq:redeom}
\rho_S(t) =  \exv{\timo  \exp\left[\int_0^t  \mathcal{L}_I(t')\, dt' \right]}_B \rho_S (0),
\end{equation}
where $\exv{\cdot}_B$ denotes an expectation value with respect to the initial state of the bath.

If we further assume that the initial state of the bath is Gaussian we can apply the relation $\langle \text{e}^{X}\rangle=\text{e}^{\langle X^2 \rangle/2+\langle X \rangle}$ which is valid for expectations taken over any $X$ that has a Gaussian distribution. This condition holds for $\mathcal{L}_I$, which is linear in the $a_q^{(\dagger)}$, when the expectation is taken with respect to 
$\rho_B$. We have defined the Hamiltonians such that $\langle \mathcal{L}_I \rangle=0$ and so the environment part of Eq.~\eqref{eq:redeom} can be simplified to:
\begin{multline} \label{eq:influence}
\exv{\timo \exp\left[\int_0^t  \mathcal{L}_I(t') \, dt' \right]}_B =\\ \exp \left[\frac{1}{2}\int_0^t  \int_0^{t} \exv{\timo \mathcal{L}_I(t')\mathcal{L}_I(t'')}_B \, dt'' \, dt' \right].
\end{multline}
The right-hand side of the above equation is exactly the Feynman-Vernon influence functional~\cite{feynman1963rp} in superoperator form. This relation holds for any $\mathcal{L}_I(t)$ that is linear in $a_q$ and $a_q^\dagger$, and so we can add extra such terms that will allow us to use a moment generating function approach to calculate expectation values of bath operators, as we will now explain. We include the extra terms by making the replacement $\mathcal{L}_I(t)\rightarrow \overline{\mathcal{L}_I}(\Lambda,\Lambda^*,t)$ with:
\begin{widetext}
\begin{equation}
\label{eq:LI}
    \overline{\mathcal{L}_I}(\Lambda,\Lambda^*,t)=\mathcal{L}_I(t)+  \sum_q\sum_{\alpha=L,R} \Big[\Lambda_q^\alpha(t) \mathcal{A}_q^\alpha (t)+   \Lambda^{\alpha*}_q  (t) \mathcal{A}_q^{\alpha\dagger} (t)\Big],
\end{equation}
where $\Lambda_q^{\alpha(*)}(t)$ are scalar valued functions associated with superoperators $\mathcal{A}_q^{\alpha(\dagger)} (t)$. These superoperators can be expressed in terms of $a^{(\dagger)}$ acting either to the left or right of $\rho$:
\begin{align}
    \mathcal{A}_q^{L(\dagger)}\rho=a_q^{(\dagger)} \rho, & &
    \mathcal{A}_q^{R(\dagger)}\rho=\rho a_q^{(\dagger)}.
\end{align}
 
Using Eq.~\eqref{eq:LI} we can define a moment generating functional
\begin{equation}
\label{eq:GLL}
G (\Lambda,\Lambda^*) = \exv{\timo \exp\left[\int_0^t  \overline{\mathcal{L}_I}(\Lambda,\Lambda^*,t')\, dt' \right]}\exp\left[\phi(\Lambda,\Lambda^*,t)\right],
\end{equation}
where the angular brackets now denote the expectation be taken with respect to the total initial density matrix $\rho$. We have included a scalar valued function $\phi$ in the definition of $G$; the exact form of $\phi$ is dependent on the order of the ladder operators in the quantity of interest. This arises due to Baker-Campbell-Hausdorff factors that appear in the construction of the generating functional, see App.~\ref{app:num} for more details. As an example any normal-ordered expression results in
\begin{align}
\phi(\Lambda,\Lambda^*,t) &=  \sum_q \sum_{\alpha=L,R}  \sum_{q'} \sum_{\alpha'=L,R}\int_0^t dt' \int_0^t dt'' \Lambda^{\alpha*}_q(t') \Lambda_{q'}^{\alpha'}(t'') [\mathcal{A}^{\alpha\dagger}_q (t'),\mathcal{A}^{\alpha'}_{q'}(t'')] \nonumber \\
&= \sum_q \int_0^t dt' \int_0^t dt'' \left( \Lambda^{R*}_q(t') \Lambda_{q}^{R}(t'') - \Lambda^{L*}_q(t') \Lambda_{q}^{L}(t'')\right)\delta(t'-t'')  \nonumber \\
&= \sum_q \int_0^t dt' \left( \Lambda^{R*}_q(t') \Lambda_{q}^{R}(t') - \Lambda^{L*}_q(t') \Lambda_{q}^{L}(t')\right).
\end{align}

The expectation of any function of left-acting ladder operators $f$ and right-acting ladder operators $g$ can then be expressed as functional derivatives of Eq.~\eqref{eq:GLL}:
\begin{multline}
\Tra[f(\{a_q (t_m),a_q^{\dagger} (t_n)\}) \rho g(\{a_q (t_o),a_q^{\dagger} (t_p)\})] = \\ f\left(\left\{\frac{\delta}{\delta \Lambda_q^L (t_m)},\frac{\delta}{\delta \Lambda_q^{L*}(t_n)}\right\}\right) g\left(\left\{\frac{\delta}{\delta \Lambda_q^R (t_o)},\frac{\delta}{\delta \Lambda_q^{R*}(t_p)}\right\}\right)G(\Lambda,\Lambda^*)\Bigg\rvert_{\Lambda,\Lambda^*=0}.
\label{eq:Trf}
\end{multline}
\end{widetext}

In Eq.~\eqref{eq:Trf} we have implicitly performed the derivative before taking either of the traces over the system or bath. These operations commute so we can either: trace over both system and environment and then take a numerical derivative (this is the counting field approach~\cite{popovic2020non,silaev2014lindblad,kilgour2019path}); or take an analytic derivative after tracing out the bath to get an expression purely in terms of system correlation functions. We use the latter approach as it allows us to calculate multiple different bath observables from a single influence functional. 

As an example, we can use the technique introduced here to derive the following expression for the occupation of a specific mode (see App.~\ref{app:num} for details):
\begin{widetext}
\begin{multline}\label{eq:occinf}
n_q (t) = n_q (0) + |g_q|^2 \int_0^t  \int_0^t  \Tra [s(t')s(t'')\rho]\times \\\times \bigg\lbrace   \cos [ \omega_q (t'-t'')]-i \sin[ \omega_q (t'-t'')]\coth (\frac{ \omega_q}{2 T}) \bigg\rbrace \, dt'' \, dt' ,
\end{multline}
\end{widetext}
where $n_q (t) = \langle a_q^\dagger (t) a_q(t)\rangle$ and $T$ is the temperature of the bath.  Expressions such as Eq.~\eqref{eq:occinf} can be obtained for any bath correlation function, using the general technique outlined above. We give a further example of how this approach can be used to calculate multi-time correlation  
functions in App.~\ref{sec:apptwomode}. These expressions will involve integrals over system correlation functions that are of the same order or lower. Any technique can be used to calculate the required system correlation functions, though naturally any approximations made in obtaining these will be reflected in the inferred bath dynamics. We include details of how $n_q (t)$ can be calculated for a simple model including just two modes in the environment in App.~\ref{sec:bench}. Further, in App.~\ref{app:g2} we show how these results can be generalised to find higher order statistics of environment modes such as $g^{(2)}$.

To calculate a single correlation function $\Tra [s_1(t')s_2(t'')\rho]$ in the integrand of Eq.~\eqref{eq:occinf} would typically require evolving $\rho$ to time $t''$, acting with $s_2$, evolving to $t'$ and finally taking the expectation of $s_1$.
To calculate the time evolution for each required $t'$ and $t''$ needed to evaluate the integral in Eq.~\eqref{eq:occinf} is a computationally expensive procedure. To address this problem we use the process tensor approach as this requires a single (albeit individually  more expensive) calculation from which any system observable can be determined with little further cost. Below, we introduce the idea of a process tensor conceptually and give an overview of how it can be calculated efficiently by contracting a tensor network.

To summarise, in this section we have described a method for finding bath correlation functions from system correlations alone. In the next section we will introduce process tensors, which are ideal for finding system correlation functions efficiently.


\section{Process Tensors \label{sec:proctens}}

\begin{figure*}[ht]
    \centering
    \includegraphics[width=\textwidth]{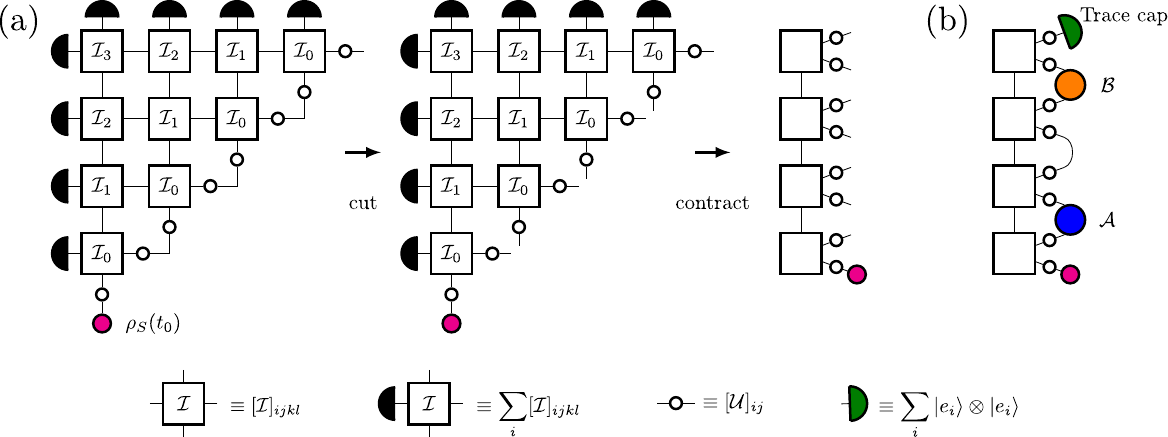}
    \caption{\label{fig:networks}(a) The leftmost network corresponds to propagation of the reduced density matrix $\rho_S$ forward 4 timesteps. Here the $\mathcal{I}$ tensors account for the non-Markovian influence of the environment and the $\mathcal{U}$ tensors the time-local system Hamiltonian evolution. Moving through the steps gives how the tensor network representation of the process tensor relates to that used for propagation. First the legs between each timestep are cut and left open and then the network is contracted column-by-column with standard tensor network routines. (b) Calculation of the correlation function $\langle \mathcal{B}(t_3)\mathcal{A}(t_1)\rangle$ using the process tensor shown in the final step of (a). For a product Hilbert space spanned by $\{\ket{e_i}\otimes \ket{e_j}\}$ the trace cap is defined as $\sum_i \ket{e_i}\otimes \ket{e_i}$, the vectorised Hilbert space identity matrix and a null eigenvector of any Liouvillian. }
    
\end{figure*}

To be able to show the full utility of the approach derived above we need to make use of a concrete method which is able to accurately provide us with the required correlation functions of system observables. To that end we employ the processes tensor formalism which we outline below. We emphasise at this point that any method capable of calculating these correlation functions can be used.

    In this section we review the process tensor formalism, which is a general operational approach to non-Markovian open quantum systems~\cite{Pollock2018}. It supposes a finite set of control operations act on the system at a sequence of time points, such as the preparation of a particular state or a projective measurement. The process tensor has two indices for each time at which a control operation can be applied. It therefore allows the computation of any multi-time correlation function of system operators by inserting control operations at the corresponding times.

    For Gaussian environments it has been shown that the process tensor can be efficiently calculated by contracting a tensor network~\cite{Jorgensen2019exploiting,OQuPy} (see also Fig.~\ref{fig:networks} which we will discuss in detail shortly). This is done by an alternative contraction order of the TEMPO network originally derived in Ref.~\cite{strathearn2017efficient} as a representation of the quasi-adiabatic path integral (QUAPI) method~\cite{makri1995tensor}.
    The process tensor fully captures the non-Markovian nature of the interaction of the system with its environment and is therefore challenging to compute. Once obtained, however, it can be used repeatedly to compute any multi-time correlation function at very little added numerical cost. With this, we have a tool at hand to evaluate the occupation, Eq.~\eqref{eq:occinf}, with moderate computational effort. In the following, and with reference to Fig.~\ref{fig:networks}, we outline the form of the TEMPO tensor network and then explain how it can be adapted to compute the process tensor. We refer the reader to~\cite{strathearn2017efficient,strathearn2020modelling} for more details on the derivation of the network and how it can be efficiently contracted.

The TEMPO network can be understood in terms of superoperators beginning from a discretisation of Eq.~\eqref{eq:redeom} with Eq.~\eqref{eq:influence} substituted:
\begin{multline}
    \rho_S(t_N) = \\ \exp \left[\sum_{k=0}^{N-1}\sum_{k'=0}^k \exv{ \mathcal{L}_{I} (t_k) \mathcal{L}_{I} (t_{k'})} \Delta t^2 \right] \rho_S(t_0) \mathrm{.}
\end{multline}
We move back into the Schr\"odinger picture and perform a symmetrised second-order Trotter splitting between the system and interaction parts:
\begin{multline}\label{eq:discprop}
\rho^{(\text{S})}_S (t_N) = \\ \timo \mathcal{U} \prod_{k=0}^{N-1} \left[ \mathcal{U}^2  \prod_{k'=0}^{k} \mathcal{I}_{k-k'}(t_k,t_{k'})\right] \mathcal{U}\rho^{(\text{S})}_S (t_0),
\end{multline}
where we 
have discretised the evolution into $N$ timesteps of length $\Delta t$ and $t_k=k\Delta t$. We have also defined the system density matrix in the Schr\"odinger picture as $\rho_S^{(\text{S})}(t)$. Here we assume $H_S$ is time independent such that $\mathcal{U}=\text{e}^{\mathcal{L}_S \Delta t/2}$, with the half-timesteps arising due to the Trotter splitting we use. For each timestep $t_k$ the terms $\prod_{k'=0}^{k} \mathcal{I}_{k-k'}(t_k,t_{k'})$ account for the influence of the past system states on $t_k$ via the environment:
\begin{equation}
	    \mathcal{I}_{k-k'}(t_k,t_{k'})=  \exp \left[\langle \mathcal{L}_{I} (t_k) \mathcal{L}_{I} (t_{k'}) \rangle \Delta t^2 \right].
	\end{equation} 
We can represent this propagation as a tensor network.  
For example, the leftmost network in Fig.~\ref{fig:networks}(a) corresponds to propagating a state forward in time by 4 steps. In the superoperator description an initial density operator in a $D$-dimensional Hilbert space is represented as a $D^2$ component vector (purple circle in Fig~\ref{fig:networks}). This can be regarded as a one-site matrix product state (MPS). Within this description the system propagators {$\cal U$} are rank-2 tensors (matrices) and the influence functions $\mathcal{I}$ are rank-4 tensors.

At each timestep the current state is contracted with a matrix product operator (MPO), corresponding to a row in the leftmost network in Fig.~\ref{fig:networks}(a), that grows the MPS by one site. In other literature the growing MPS is referred to as the augmented density tensor (ADT) and contains information about both the current system state and its correlations with the environment. The state can be extracted at each timestep by summing over all but the rightmost legs of the ADT. For example, the fully contracted network at the left in Fig.~\ref{fig:networks}(a) would be the state after four timesteps. We show how this network can be converted into a process tensor following the steps in Fig.~\ref{fig:networks}(a): The process tensor can be identified as the full propagation network with the $\mathcal{U}$ propagators between each timestep disconnected and the legs left open. This network can then be contracted from left to right and the process tensor stored as an MPO. 

    Alternatively, one can view the system propagators $\mathcal{U}$ as control operations and exclude them from the process tensor. Such a division can be exploited to efficiently explore the dynamics for different system Hamiltonians~\cite{fux2021efficient}. In this work, however, we include the system propagators in the process tensor, and we are then able to insert arbitrary operations between the open legs (for example the operations in Fig.\ref{fig:networks}(b) shown as orange and blue circles). This process allows us to efficiently calculate multitime correlation functions. Alternatively, we can retrieve the system evolution by inserting identity operations.


Carrying out the contraction step in Fig.~\ref{fig:networks}(a), column-by-column, would result in an object that grows exponentially with each column contracted. Many of the degrees of freedom of this object, however, contribute negligibly and can be discarded without significantly sacrificing accuracy~\cite{orus2014practical,schollwock2011density}. These irrelevant degrees of freedom can be systematically identified by performing a singular value decomposition (SVD) on each tensor in the chain and dispensing with any components that have singular values below a threshold value. It is difficult to make any concrete statements on what the discarded components correspond to. Regardless of this we can be confident that we have retained enough singular values by ensuring the results are converged with increasing the SVD threshold value. These SVD sweeps are carried out once in each direction after contracting a column. For all of the results in this paper we discarded components with singular values of relative magnitude less than $10^{-8}$. The other main approximation we make is a finite memory cutoff where the influence of the bath is only stored for $K$ timesteps into the past~\cite{makri1995tensor,makri1995tensor-2}. The value of $K$ depends on $\Delta t$: for convergence we require that $K$ be increased until $K\Delta t$ exceeds the correlation time of the bath. However $\Delta t$ must in turn be reduced to minimise the error from the Trotter splitting. For all the results presented below in Sec.~\ref{sec:SBM} the parameters $\Omega \Delta t=0.05$ and $K=50$ were sufficient for convergence.  These parameter values lead to more intensive calculations than would typically required for this form of coupling. This is because the calculation of bath dynamics is highly sensitive to small errors in system expectation values. 
Since these are integrated over, minor errors which propagate in time can lead to large errors in the resulting bath observables.

The final process tensor is of the form shown in the final step of Fig.~\ref{fig:networks}(a) which for $N$ time points is $N-1$ sites long; the bond dimension is related to the number of relevant bath degrees of freedom \cite{cygorek2021numerically}. 

From this a correlation such as $\langle \mathcal{B}(t_3)\mathcal{A}(t_1)\rangle$, where $\mathcal{A}$ is an arbitrary system superoperator, can be calculated by contracting the open legs of the process tensor with $\mathcal{A}$ and $\mathcal{B}$ inserted at the corresponding timesteps as shown in Fig.~\ref{fig:networks}(b). This process is straightforward to generalize to higher order correlation functions, through the contraction of further system superoperators.

To summarize this section: we have shown how the process tensor is ideally suited to finding system correlation functions of an open system that is strongly coupled to its environment. In particular, the process tensor needs only to be calculated once, before contracting with relevant system operators enable the calculation of any desired correlation function (up to the final time chosen for the process tensor). Multi-time bath correlation functions can then be found from these system correlation functions by using the method described in Sec.~\ref{sec:BD}. In the next section we will give an example of this procedure for a well known model.

\section{Biased Spin Boson Model \label{sec:SBM}}

In this section we study the simple example system of a driven spin-1/2 particle coupled to a continuum of bosonic modes, the biased spin-boson model~\cite{leggett1987dynamics,weiss2012quantum}. Studies of the spin-boson model are ubiquitous in the field of open quantum systems for two reasons. First, the Hamiltonian is relatively simple while still allowing non-trivial behaviour, making the spin-boson model inviting for those looking to test new methods. Second, it provides the paradigmatic model for dissipative two-level systems meaning it can be applied to a variety of physical processes for example quantum dot emission in the presence of phonons~\cite{nazir2016, strathearn2020modelling}, energy transfer in biological or molecular systems~\cite{rackovsky1973electronic,rey2013exploiting,gribben2020exact,iles2016energy}, the interaction of superconducting qubits with microwave waveguides~\cite{zheng13} quantum dots interacting with micromechanical resonators~\cite{yeo13} and energy transport in solid state systems~\cite{rozbicki2008quantum}.  
The Hamiltonian of the system is:
\begin{equation}
H = \epsilon s_z + \Omega s_x + s_z \sum_q g_q (a_q + a_q^\dagger) + \sum_q \omega_q a_q^\dagger a_q,
\end{equation}
where we have introduced the spin-1/2 operators $s_z=(\dyad{1}{1}-\dyad{0}{0})/2$ and $s_x=(\dyad{1}{0}+\dyad{0}{1})/2$ which act on the states $\{\ket{0},\ket{1}\}$. The transition between these states is driven classically with strength $\Omega$ and bias $\epsilon$. The system is in turn coupled with strength $g_q$ to a bath of bosonic modes of energy $\omega_q$. The bosonic degrees of freedom are described by ladder operators $a_q$ and $a_q^\dagger$ which satisfy the canonical commutation relations: $[a_q,a_{q'}^\dagger]=\delta_{qq'}$ and $[a_q,a_{q'}]=[a_q^\dagger,a_{q'}^\dagger]=0$.

The environment in this model is characterised by its spectral density which captures both the coupling to each mode $g_q$ and the density of states of the environment. It is defined as:
\begin{equation}
J(\omega) = \sum_q |g_q|^2 \delta (\omega - \omega_q).
\end{equation}
This can be related to the auto-correlation function of the bath coupling operator $B= \sum_q g_q (a_q + a_q^\dagger)$,
\begin{align}
    C(t)= \exv{B(t)B(0)}  = \int_0^\infty C (\omega,t) \, d\omega , 
\end{align}
where we have introduced 
\begin{equation}
    C(\omega,t)=J(\omega) \bigg[\cos(\omega t) \coth\left(\frac{\omega}{2 T}\right)   -i \sin (\omega t)\bigg]
\end{equation}
and $T$ is the temperature of the bath. We will also find it useful to define the derivative of this quantity $\tilde{C}(\omega, t) = \partial C(\omega, t)/\partial t$.

The spectral density is typically given a functional form informed by experimental measurements. Here we use an Ohmic spectral density with an exponential cutoff~\cite{leggett1987dynamics}:
\begin{equation}
J(\omega) = 2 \alpha \omega \text{e}^{-\omega/\omega_c},
\end{equation}
where $\omega_c$ is the cut-off frequency and $\alpha$ a dimensionless coupling constant.

\subsection{Steady State Properties}

To analyze the behaviour of the system by looking at both the system and environment degrees of freedom we wish to look at how energy is redistributed during the dynamics induced by the spin boson model. To this end we consider the change in energy of each bath mode, defined as:
\begin{equation}
\Delta Q_q(t) = \omega_q \left[\langle a_q^\dagger (t) a_q(t)\rangle -\langle a_q^\dagger (0) a_q (0)\rangle \right].
\end{equation}
This quantity gives the difference in energy of a particular mode $q$ compared to its initial, thermal equilibrium, value. 
However, this is ill-defined in the continuum limit. We cannot refer to the energy change of a specific mode of the environment as the coupling to any single mode is infinitesimal. Instead we consider the change over a narrow band of modes: 
\begin{widetext}
\begin{equation}\label{eq:occ}
\Delta Q (\omega,t) =   i \int_0^t  \int_0^t  \Tra [s(t')s(t'')\rho] \int_{\omega-\delta/2}^{\omega+\delta/2}  \tilde{C} (\omega',t'-t'') \, d\omega' \, dt'' \, dt'.
\end{equation}
\end{widetext}
The frequency interval is of width $\delta$ centered at $\omega$. We can then use the machinery described above to compute this quantity in the steady state of the spin boson model.

For the results in this section we choose a relatively weak system-bath coupling. This is not due to any inherent limitations of the methods used but rather because the behaviour in this limit is more interesting. At stronger coupling contributions from the re-organisation energy of the environment dominate and the behavior reduces to that of the exactly solvable independent boson model. This effect has been seen in previous calculations involving the total heat of the bath~\cite{popovic2020non}. 

\begin{figure}[t!]
\ig{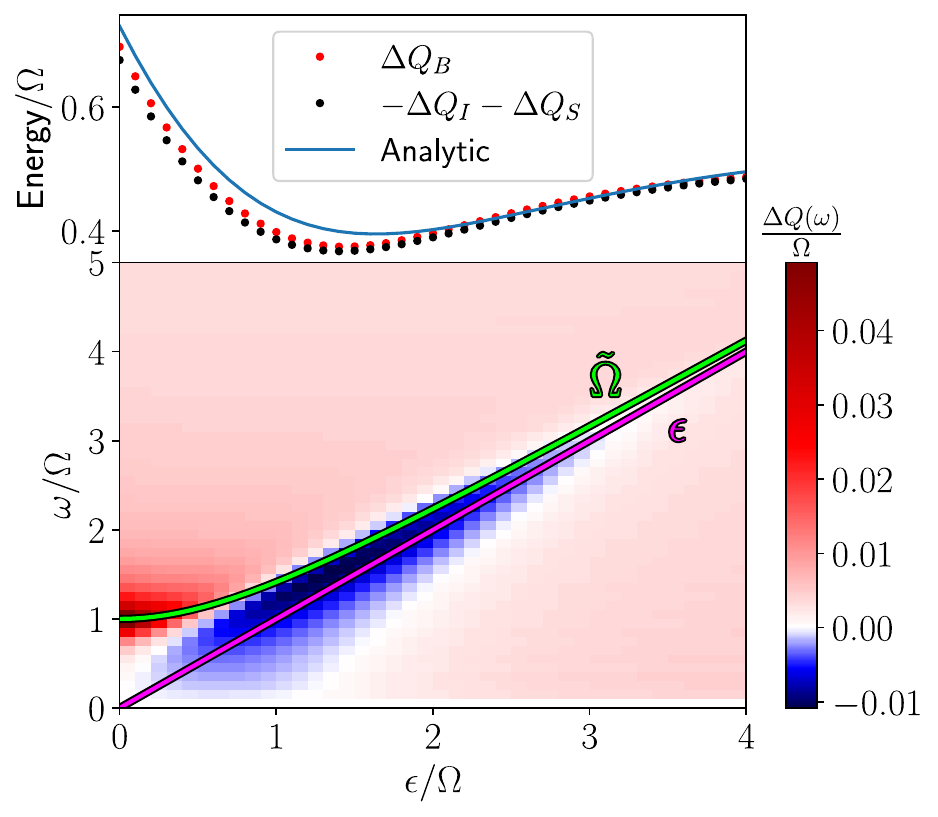}
\caption{\label{fig:varybias} Numerically exact frequency resolved steady-state variation of the heat in bath modes as a function of bias $\epsilon$ (main panel) and the total heat exchanged with the bath compared with the analytic expression given by Eq.~\eqref{eq:totbathapp} (upper panel). In the main panel both the bias $\epsilon$ and Rabi frequency $\tilde{\Omega}=\sqrt{\Omega^2+\epsilon^2}$ are overlaid for comparison. In the upper panel we also plot the total change in system and interaction energy $-\Delta Q_S-\Delta Q_I$ as a check that our result satisfies energy conservation i.e.~$\Delta Q_B = -\Delta Q_I -\Delta Q_S $. Other parameters are $T=\Omega$, $\alpha=0.05$, $\omega_c=10\Omega$ and $\delta=0.1\Omega$. For all results the initial state is $\ket{0}$.}
\end{figure}

\begin{figure*}[t!]
\ig{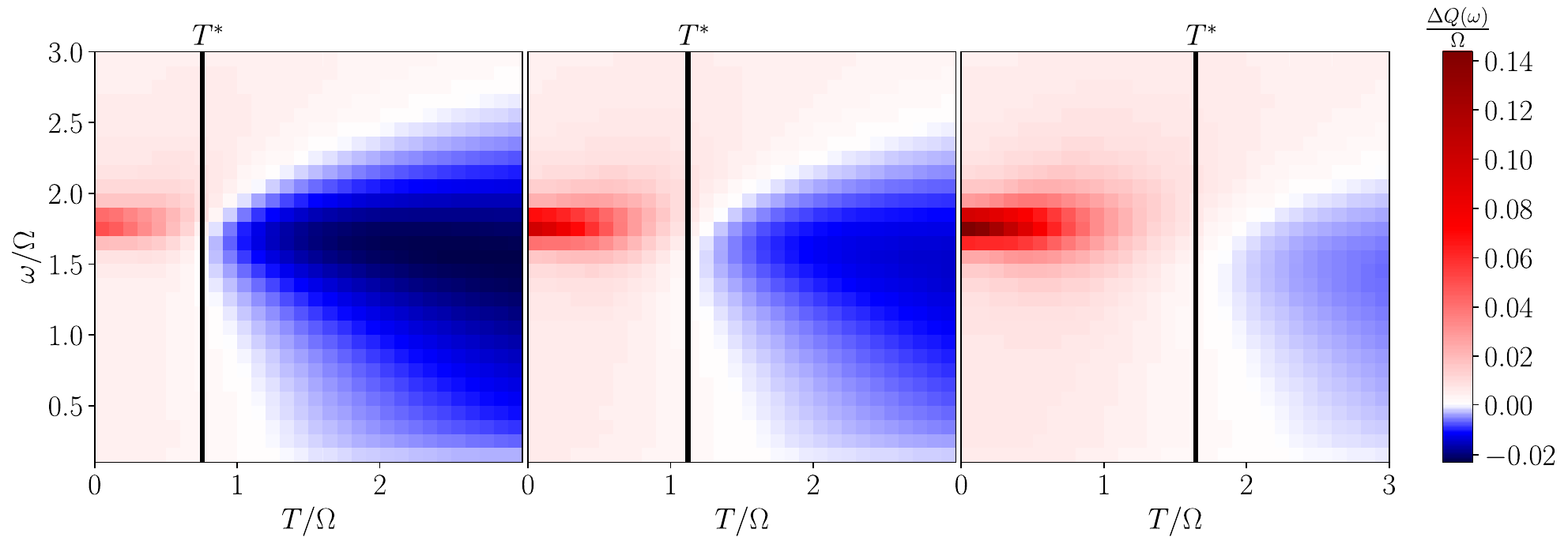}
\caption{\label{fig:crossover} Steady state change in bath heat, $\Delta Q_B$, as a function of temperature for varying initial state where  (a) $p=1.0$, (b) $p=0.9$ and (c) $p=0.8$. For each plot we have marked the crossover temperature $T^*$ as predicted by Eq.~\eqref{eq:crossoverT} with a vertical black line. Here we fixed $\epsilon=1.5\Omega$ and all other parameters are unchanged from Figure \ref{fig:varybias}.}
\end{figure*}

In the main panel of Fig.~\ref{fig:varybias} we show the period-averaged steady state value for $\Delta Q$ as a function of the bias $\epsilon$ with the two-level system initialised in the low energy state, $\ket{0}$. This initial state is uncorrelated with the environment and so as the system moves towards thermal equilibrium energy needs to flow between the system and environment: it is this that we track.  The period-averaging is performed over a single oscillation of each mode, such that we calculate
\begin{equation}
\Delta Q (\omega) = \frac{\omega}{2\pi} \int_{t_{SS}}^{t_{SS}+2 \pi/\omega}  \Delta Q (\omega,t') \, dt' ,
\end{equation}
where $t_{SS}$ is chosen to be long enough that the system density operator is stationary $\dot{\rho}_S (t~\geq~t_{SS}) \simeq 0$. We do this because even after the system degrees of freedom reach a stationary state there are still oscillations in the mode occupations which need to be averaged over to find the appropriate quantity. 
This change in energy is then able to probe the way in which energy is removed from or added to particular environment modes to populate the spin-1/2 in its stationary configuration. We see that for small bias the initial system energy is higher than that in thermal equilibrium and hence the environment gains energy from the system, while at large bias the opposite is true. This might seem at odds with our later analysis which shows that the total bath energy increases at all values of $\epsilon$. This is a result of the energy gained from the interaction being switched on at $t=0$, known as the re-organisation energy. The contribution from the re-organisation energy appears in the main panel of Fig.~\ref{fig:varybias} as a pale red background heating. It is only through the mode resolution of our approach that we see that it is possible for the direction of energy flow into/out of the bath to be frequency-dependent. In the following we will expand on the intuition for these observed effects.

To confirm the accuracy of this approach we also checked that the total energy was conserved. This required calculation of the total change in $\langle H_S \rangle$, $\langle H_I \rangle$ and $\langle H_B \rangle$ which we will denote $\Delta Q_S$, $\Delta Q_I$ and $\Delta Q_B$ respectively. The change in system energy, $\Delta Q_S$ is trivial to calculate from the dynamics. The total energy exchanged with the bath, $\Delta Q_B$, is given by
\begin{multline}\label{eq:totbath}
\Delta Q_B = i\int_0^{t_{SS}} \int_0^{t_{SS}}   \Tra [s(t')s(t'')\rho] \\ \times \tilde{C} (t'-t'') \, dt'' \, dt' .
\end{multline}
The bath will generally have a finite memory time $t_K$ such that the auto-correlation function satisfies $C(t>t_K)\simeq0$ and hence the derivative also vanishes. Therefore if the total energy exchanged is the relevant quantity then only correlation functions that span this memory time are required.

A similar expression can be derived for $\Delta Q_I$ again using the generating functional formalism. This is given by:
\begin{multline}\label{eq:totint}
\Delta Q_I = 2 \Im  \left[ \int_0^{t_{SS}}    \Tra [s(t_{SS})s(t')\rho] \right. \\ \left. \times C(t-t') \, dt' \vphantom{\int_0^{t_{SS}}} \right]. 
\end{multline}

All of the bath observables calculated here can be derived from the same set of correlation functions.  The only variation is in the form of the kernel used for the integral transformation. Generally the dynamics of an $n^{\text{th}}$ order bath correlation function up to time $t$ depends on the set of $n^{\text{th}}$ order system correlation functions up to that point.

In the upper panel of Fig.~\ref{fig:varybias} we compare the numerical calculation of the heat gained by the bath, $\Delta Q_B$ with the remaining contribution $-(\Delta Q_I+\Delta Q_S)$. We see that, up to small numerical inaccuracies, these quantities are effectively equal as would be expected from energy conservation.

To gain a better understanding of the results seen here we proceed to develop semi-analytic expressions for the changes in energy observed. In Ref.~\cite{popovic2020non} it was shown that for weak coupling $\Delta Q_B$ can be well approximated by
\begin{equation}\label{eq:totbathapp1}
    \Delta Q_B \approx E_r-\Delta Q_S^{G},
\end{equation}
where $E_r$ is the reorganisation energy of the bath given by 
\begin{equation}\label{eq:reorg}
    E_r = \frac{1}{2}\int_0^\infty  \frac{J(\omega')}{\omega'}d\omega'=\alpha \omega_c,
\end{equation}
and $\Delta Q_S^G$ is approximate change in system energy defined as
\begin{equation}
    \Delta Q_S^G = \Tra\{H_S[\rho_{SS}(T)-\rho_S(0)]\},
\end{equation}
where $\rho_{SS}(T)$ is the reduced Gibbs state of the system Hamiltonian,
\begin{equation}
\rho_S (t_{SS}) \approx \rho_{SS} (T) = \frac{\text{e}^{-H_S/T}}{\mathrm{Tr} [\text{e}^{-H_S/T}]}.
\end{equation}
Since the coupling to the environment is weak we expect there to be no significant correlations between the system and environment such that the system simply reaches a state in approximate thermal equilibrium with respect to $H_S$ at the temperature of the environment.

From this we can then find an analytic expression for $\Delta Q_S^G$:
\begin{equation}\label{eq:systemheat}
\Delta Q_S^G= \frac{1}{2} \left(\epsilon-\tilde{\Omega} \tanh
   \left(\frac{\tilde{\Omega}}{2 T}\right)\right),
\end{equation}
where $\tilde{\Omega}=\sqrt{\Omega^2+\epsilon^2}$ is the generalised Rabi frequency and we have used $\rho_S(0)=\ket{0}\bra{0}$. This in turn allows us to derive an (approximate) analytic expression for $\Delta Q_B$ by substituting~\eqref{eq:systemheat} and~\eqref{eq:reorg} into~\eqref{eq:totbathapp1} giving
\begin{equation}\label{eq:totbathapp}
    \Delta Q_B \approx \frac{1}{2} \left( 2\alpha \omega_c - \epsilon+\tilde{\Omega} \tanh
   \left(\frac{\tilde{\Omega}}{2 T}\right)\right).
\end{equation}
In the upper panel of Fig.~\ref{fig:varybias} we compare the analytic approximation given by Eq.~\eqref{eq:totbathapp} with the numerical results and find good agreement as would be expected at weak system environment coupling.


From the form of Eq~\eqref{eq:systemheat} we can identify a crossover temperature for the environment, $T^*$, above which the dynamics induced by the coupling to the environment increases the system energy and vice-versa. This is given by the solution of
\begin{equation}
\Tra_S\{H_S[\rho_{SS}(T^*)-\rho_S(0)]\}=0.
\end{equation}
For initial states that are incoherent mixtures of the two basis states $\rho_S(0)=p\ket{0}\bra{0} + (1-p)\ket{1}\bra{1}$ we can solve this to find
\begin{equation}\label{eq:crossoverT}
T^* = \frac{\tilde{\Omega}}{2 \tanh^{-1} \left[\epsilon(2 p-1)/\tilde{\Omega}  \right]}.
\end{equation}
In Fig.~\ref{fig:crossover} we show the steady state heat change in the environment $\Delta Q(\omega)$ as a function of temperature for three values of $p$. In each case we see that for temperatures below $T^*$ the bath absorbs energy from the system and at temperatures above $T^*$ the opposite occurs. The main transfer occurs at a frequency close to the Rabi frequency of the system. However, as temperature increases the heat transfer occurs across a broader distribution of modes. This is linked to the broadening of the Bose-Einstein distribution meaning that the system can absorb energy from a wider range of frequencies. From this we can see that the analytic expression is a good predictor of whether the system has absorbed or emitted energy into the bath at a specific frequency.

\begin{figure}[t!]
\ig{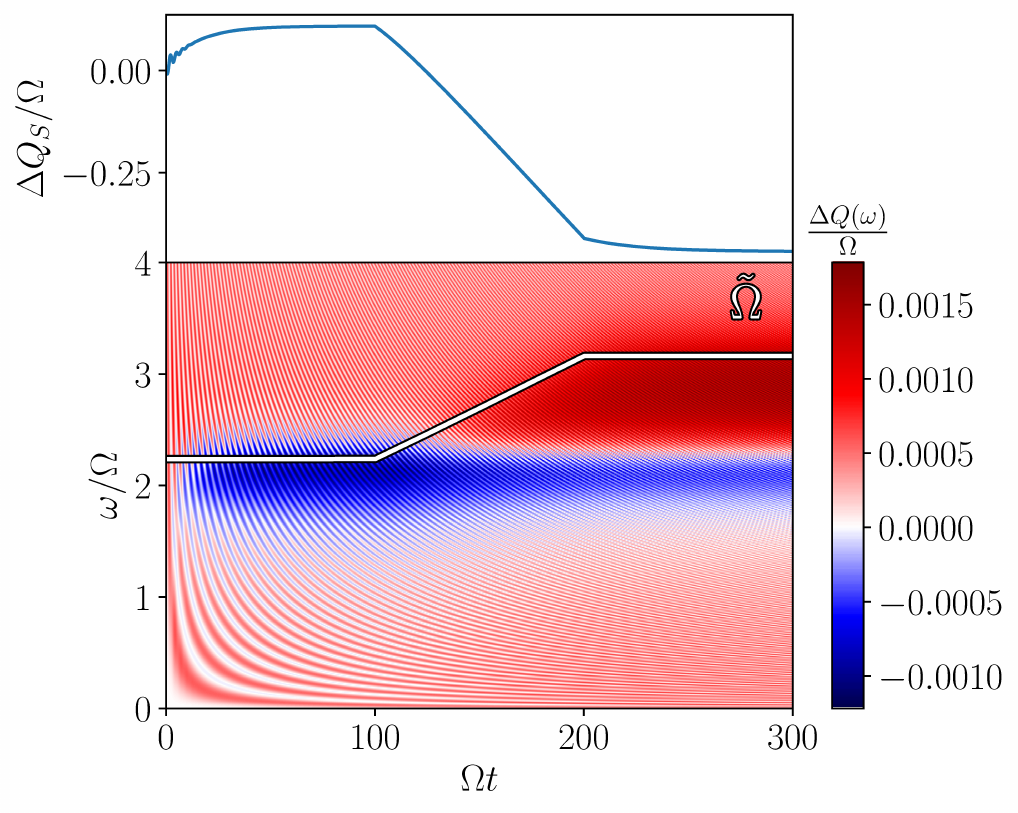}
\caption{\label{fig:drivingswitch} Movement of heat around the bath over time using the driving routine described in Eq.~\eqref{eq:routine} with $\Omega t_1=100$, $\epsilon_1=2\Omega$, $\Omega t_2 = 200$, $\epsilon_2=3\Omega$ and $\delta=0.01\Omega$. Here the generalised Rabi frequency $\tilde{\Omega}=\sqrt{\Omega^2+\epsilon^2}$ is overlaid for comparison. The other parameters are the same as Fig.~\ref{fig:varybias}. The upper panel shows the system energy.}
\end{figure}

\subsection{Dynamics}

We now move on to show how the methods described above can also be used to access the complete dynamics of the bath. To do this we implement a time-dependent driving sequence with the goal of demonstrating the absorption of energy from the bath over a range of frequencies followed by emission into the bath over a different frequency  range. This shows how, by controlling the dynamics of the system, we can control the state of the environment. We initialize  the system in the $\ket{0}$ state at $t=0$ and allow the bias to vary over time with the system Hamiltonian
\begin{equation}
    H_S(t)=\epsilon(t) s_z + \Omega s_x.
\end{equation}
We use a linear ramp for the bias and hence set
\begin{equation}\label{eq:routine}
\epsilon(t) = 
\begin{cases}
\epsilon_1 & 0 \leq t<t_1 \\
\epsilon_1+(\epsilon_2-\epsilon_1) \frac{t-t_1}{t_2-t_1} & t_1 \leq t \leq t_2\\
\epsilon_2 & t > t_2
\end{cases}.
\end{equation}
For times $t<t_1$ this is identical to the problem studied above. Hence, we choose $t_1$ to be large enough to allow the system to first reach thermal equilibrium with the environment and $\epsilon_1$ is chosen such that this equilibrium state is higher in energy than the initial state and thus absorbs energy from the bath. The bias is then linearly increased, increasing the system energy along with it. The system is then in a state such that energy must be emitted into the bath for the system to be in equilibrium with respect to the new system Hamiltonian, which now has a much higher value for $\epsilon$. This behaviour is shown in Fig.~\ref{fig:drivingswitch}. As can be seen in the upper panel of the Figure 
during the linear ramp the energy of the system decreases, emitting energy into the bath. This can be seen in the heat distribution of the bath as a band of modes which increase in energy. The modes targeted correspond to frequencies close to the (time-dependent) Rabi frequency $\tilde{\Omega}(t)=\sqrt{\epsilon(t)^2+\Omega^2}$ which changes over the course of the linear ramp. This shows that by using a time varying drive, along with the intuition we have built up about the behaviour of the environment, we are able to control the (frequency dependent) absorption and emission of energy into and out of the bath.

The spin-boson model has provided a concrete example of how we can calculate the dynamics of bath modes, by combining the moment generating functional method introduced in Sec.~\ref{sec:BD} and the process tensor approach of Sec.~\ref{sec:proctens}. We saw how the steady state behaviour can be calculated as well as how we can control the population of different modes by using an appropriate time-dependent drive.

\section{Conclusions \& Outlook}

In this paper we have introduced a method for calculating any dynamical observable of a Gaussian bath purely in terms of correlation functions of a linearly coupled system. 
For systems coupled to infinite baths we demonstrated that the process tensor formalism provides a framework to calculate, cheaply, the large number of system correlation functions needed. In general it is numerically costly to calculate a process tensor. However, once it exists it can then be used to completely characterise any time dependent observable or any multi-point correlation function of the system, bath, or a combination of both. 

Specifically, we were able to calculate how energy is moved between different environment modes and a spin-1/2 particle in the biased spin boson model. We showed that, for weak system environment coupling, the spin reaches thermal equilibrium with the bath removing or adding energy to modes which correspond to the system frequency. We also saw how the techniques described here can allow us to track the flow of energy into and away from the system when it is subject to a time dependent driving. 

The framework we presented is not specific to any technique for calculating the required correlation functions and opens up many avenues to analyze a broad range of open quantum systems. This includes simulations of any Gaussian environment, including, for example, fermionic environments. For non-Gaussian environments, the moment generating functional approach is still valid, but Wick's theorem does not hold so the derivatives in Eq.~\ref{eq:GLL} must to carried out numerically. 

Recent advances have extended the applicability of numerically exact process tensor techniques to many-body systems~\cite{fux2022thermalization} and multiple environments~\cite{gribben2022exact}. Those techniques could be combined with the moment generating functional method we have described here to expose the dynamics of environments in such situations.

When the environment induces complex non-Markovian dynamics in a system this can lead to advantageous effects, e.g. in increased quantum channel capacities~\cite{bylicka2014non}, entanglement distribution~\cite{xiang2014entanglement} and quantum control~\cite{reich2015exploiting}. Our method has the potential to reveal the origin of such complicated non-Markovian dynamical effects, and to aid the design of new systems that could control and harness these. During the preparation of this work we have implemented these methods into an open source package~\cite{OQuPy} in an effort to enable others to easily apply these results in their own fields.


\begin{acknowledgements}
DG and GF acknowledge studentship funding from EPSRC (EP/L015110/1). AS acknowledges support from the Australian Research Council Centre of Excellence for Engineered
Quantum Systems (EQUS, CE170100009). We acknowledge support from EPSRC (EP/T014032/1). This work used EPCC's Cirrus HPC Service
(https://www.epcc.ed.ac.uk/cirrus).
\end{acknowledgements}


\printbibliography

\widetext
\clearpage
\appendix

\section{ Superoperator formalism \label{app:sup}}
To calculate the evolution of the total system-bath density matrix we start with the von Neumann equation (in the interaction picture) and use the superoperator formalism:
\begin{equation}
\frac{d}{dt}\rho(t) = -i [H_I(t),\rho(t)]  \rightarrow  \frac{d}{dt}\sket{\rho(t)} = \mathcal{L}_I(t)\sket{\rho(t)}
\end{equation}
where $\mathcal{L}_I(t)$ is the superoperator associated with taking the commutator with $-i H_I(t)$ i.e. $\mathcal{L}_I(t) \ket{\rho(t)}\rangle =-i\ket{[H_I(t),\rho(t)]}\rangle$. We can then formally integrate this and arrive at the expression:
\begin{equation}
\sket{\rho(t)}= \timo  \text{e}^{\int_0^t dt' \mathcal{L}_I(t')} \sket{\rho(0)}.
\end{equation}
To find the dynamics of reduced system we trace over the bath to leave
\begin{equation}
\sket{\rho_S(t)} = \exv{\timo  \text{e}^{\int_0^t dt' \mathcal{L}_I(t')}}_B \sket{\rho_S (0)},
\end{equation}
where we have made the assumption of a separable initial state such that $\sket{\rho(0)}=\sket{\rho_S(0)}\otimes\sket{\rho_B(0)}$ and $\langle \cdot \rangle_B$ represents an expectation taken with respect to $\sket{\rho_B(0)}$.

Throughout this paper we represent operators in Hilbert space as lowercase letters in normal italics (e.g.~$s$ and $x$) and superoperators in Liouville space superoperators as uppercase calligraphic letters (e.g.~$\mathcal{L}$). If a superoperator corresponds to simple extension of an operator then it will be labelled as such, for example:
\begin{equation}
[a,\rho] = a \rho - \rho a \rightarrow \mathcal{A}^- \sket{\rho}.
\end{equation}
In a similar vein we can define the left-acting superoperator $a\rho\rightarrow \mathcal{A}^L\sket{\rho}$, the right-acting superoperator $\rho a\rightarrow \mathcal{A}^R\sket{\rho}$ and the anti-commutator superoperator $\{a,\rho\} \rightarrow \mathcal{A}^+ \sket{\rho}$.
The only exception is that when referring to superoperators which correspond to Hamiltonians in Hilbert space we will use the typical $\mathcal{L}$ symbol. Although the time-ordering is carried out on superoperators we reserve the calligraphic $\mathcal{T}$ for the trace operation.

In Hilbert space the trace of an operator returns a scalar. In Liouville space the density matrix is vectorised and thus to return a scalar the trace must correspond to an inner product. To see how we define this we give a brief overview of the formalism of superoperators. 

Consider a Hilbert space spanned by the vectors $\{\ket{e_i}\}$ a density matrix can be written as a vector in the corresponding Liouville space by using the mapping
\begin{equation}
\rho = \sum_{i,j} \rho_{ij} \dyad{e_i}{e_j} \rightarrow \sket{\rho} = \sum_{i,j} \rho_{ij} \ket{e_i}\otimes \ket{e_j}.
\end{equation}
An equivalent mapping can be carried out on any operator in Hilbert space. We can then define the trace as a dual vector in the Liouville space which acts as follows
\begin{equation}
\Tra [\rho] = \sum_k \sum_{i,j} \rho_{ij} \braket{ e_k \dyad{e_i}{e_j}  e_k } \rightarrow \sbraket{\mathcal{T} | \rho} = \sum_k \sum_{i,j} \rho_{ij} \braket{e_k | e_i}\otimes \braket{e_k | e_j}.
\end{equation}
From this we can see that the correct dual vector is
\begin{equation}
\sbra{\mathcal{T}} \equiv \sum_k \bra{e_k}\otimes \bra{e_k}.
\end{equation}

Now consider a bipartite system $\mathcal{H}_A \otimes \mathcal{H}_B$ spanned by $\{ \ket{f_i} \otimes \ket{h_k} \}$. The corresponding Liouville space is in turn spanned by $\{\ket{f_i} \otimes \ket{f_j}\otimes\ket{h_k} \otimes \ket{h_l}\}$. The partial trace over $\mathcal{H}_B$ is then
\begin{equation}
\mathrm{Tr}_B [\cdot ] \rightarrow \sbra{\mathcal{T}_B} \equiv \sum_{i,j,k} \dyad{f_i}{f_i} \otimes \dyad{f_j}{f_j}\otimes\bra{h_k} \otimes \bra{h_k}.
\end{equation}
We now define the partial trace over a series of superoperators as:
\begin{equation}
\sbra{\mathcal{T}_B}  \mathcal{A}(t_A) \mathcal{B}(t_B) \mathcal{C}(t_C) \dots \sket{\rho_B} \equiv \braket{\mathcal{A}(t_A) \mathcal{B}(t_B) \mathcal{C}(t_C) \dots}_B.
\end{equation}
We will reserve the use of angular brackets to refer to expectations of superoperators. If there is no subscript after the angular brackets then the trace is over the entire space.

If we consider the bath to initially be in a Gaussian state, e.g. thermal equilibrium, then the trace over the bath can be simplified using Wick's theorem to give:
\begin{equation}\label{eq:wick}
 \exv{\timo\text{e}^{\int_0^t dt' \mathcal{L}_I(t')}}_B = \exp \left(\frac{1}{2}\int_0^t dt' \int_0^{t} dt'' \exv{\timo \mathcal{L}_I(t')\mathcal{L}_I(t'')}_B \right).
\end{equation}
The key result of this paper relies on an equivalent relation holding when source terms linear in the ladder operators are added to the exponent.

\section{Number operator derivation \label{app:num}}

In this section we derive Eq.~\eqref{eq:occinf} of the main text. For simplicity we consider a bath that consists of just a single mode such that $H_I=gs(a^\dagger+a)$ and $H_B=\omega a^\dagger a$, where $s$ is an arbitrary system operator. The generalisation to a multimode bath is straightforward. We wish to calculate $n(t)=\Tra [a^\dagger(t) a(t) \rho]$ and write this quantity in terms of expectation values of superoperators as follows,
\begin{equation}\label{eq:nsop}
    n(t)=\exv{\timo  \mathcal{A}^{L \dagger}(t) \mathcal{A}^L (t)  \text{e}^{\int_0^t \mathcal{L}_I (t') dt'}}.
\end{equation}
Now we can write
\begin{align}\label{eq:enterders}
  \mathcal{A}^{L \dagger}(t) \mathcal{A}^L (t) &= \left.\frac{d^2}{d \Lambda d \Lambda^*}\text{e}^{ \Lambda^* \mathcal{A}^{L \dagger}(t)} \text{e}^{ \Lambda \mathcal{A}^{L}(t)} \right|_{ \Lambda, \Lambda^*=0} \nonumber \\
  &=\left. \frac{d^2}{d \Lambda d \Lambda^*} \text{e}^{\Lambda^* \mathcal{A}^{L \dagger}(t)+ \Lambda \mathcal{A}^{L}(t)- \frac{ |\Lambda|^2}{2}} \right|_{ \Lambda, \Lambda^*=0},
\end{align}
where we have combined the exponentials in the second line using the Baker-Campbell-Hausdorff (BCH) formula and $[\mathcal{A}^{L \dagger}(t), \mathcal{A}^L (t)]=[a^\dagger(t),a(t)]=-1$. Care must be taken when explicitly calculating the derivatives above since the operators involved do not commute.

Inserting Eq.~\eqref{eq:enterders} into Eq.~\eqref{eq:nsop} gives
\begin{equation}\label{eq:ngen}
    n(t)=\left.\frac{d^2}{d \Lambda d \Lambda^*}\exv{\timo  \text{e}^{ \Lambda \mathcal{A}^{L}(t)+\Lambda^* \mathcal{A}^{L \dagger}(t)- \frac{ |\Lambda|^2}{2}}  \text{e}^{\int_0^t \mathcal{L}_I (t') dt'}}\right|_{ \Lambda, \Lambda^*=0}.
\end{equation}
We can combine the exponentials in Eq.~\eqref{eq:ngen} without using the BCH formula by virtue of the identity
\begin{equation}
    \timo \left[\mathcal{B}(t),\int \mathcal{C}(t')dt'\right]=0,
\end{equation}
which holds for arbitrary $\mathcal{B}$, $\mathcal{C}$, and integral limits. This is because the portion of the integral that contains $\mathcal{C}(t)$, which in general does not commute with $\mathcal{B}(t)$ under time ordering, is infinitesimal, i.e.~has measure zero. Thus, Eq.~\eqref{eq:ngen} becomes
\begin{align}\label{eq:ngen2}
    n(t)=\left.\frac{d^2}{d \Lambda d \Lambda^*}\exv{\timo   \text{e}^{\int_0^t \overline{\mathcal{L}_I} (\Lambda,\Lambda^*,t') dt'}}\right|_{ \Lambda, \Lambda^*=0} 
    =\left.\frac{d^2}{d \Lambda d \Lambda^*}G(\Lambda,\Lambda^*)\right|_{ \Lambda, \Lambda^*=0},
\end{align}
with
\begin{equation}
    \overline{\mathcal{L}_I} (\Lambda,\Lambda^*,t')= \mathcal{L}_I (t') +\delta(t-t')\left(\Lambda \mathcal{A}^{L}(t')+\Lambda^* \mathcal{A}^{L \dagger}(t')- \frac{ |\Lambda|^2}{2}\right).
\end{equation}
To evaluate $G(\Lambda,\Lambda^*)$ we now split the trace over the full system in Eq.~\eqref{eq:ngen2} into a partial trace over the bath, followed by a partial trace over the system
\begin{equation}
 G(\Lambda,\Lambda^*)=\exv{\timo \exv{\timo   \text{e}^{\int_0^t \overline{\mathcal{L}_I} (\Lambda,\Lambda^*,t') dt'}}_B  }_S,
\end{equation}
where we have used idempotency of time ordering, $\timo=\timo\timo$. Since the operational part of $\overline{\mathcal{L}_I}(\Lambda,\Lambda^*,t')$ is linear in bath operators we can use standard Gaussian integral results to evaluate the trace over the bath analytically. This gives
\begin{multline}\label{eq:bathtraced}
 \exv{\timo   \text{e}^{\int_0^t \overline{\mathcal{L}_I} (\Lambda,\Lambda^*,t') dt'}}_B  \\ = \exp\left(\Phi (t) + \Lambda^* \int_0^t dt' \exv{  \mathcal{A}^{L \dagger} (t)\mathcal{L}_I(t')}_B +\Lambda \int_0^t dt' \exv{  \mathcal{A}^{L} (t)\mathcal{L}_I(t')}_B+ |\Lambda|^2 n(0) \right),   
\end{multline}
where $\Phi (t)=\frac{1}{2}\int_0^t dt' \int_0^{t} dt'' \exv{\timo \mathcal{L}_I(t')\mathcal{L}_I(t'')}_B $ is the usual Feynman-Vernon influence phase (i.e.~the exponent in Eq.~\eqref{eq:wick}) and  $n(0)=\Tra[ a^\dagger(0) a(0)\rho ]$. Since this is the only part of $G(\Lambda,\Lambda^*)$ that carries dependence on $\Lambda$ and $\Lambda^*$ we can now take the derivatives and set $\Lambda=\Lambda^*=0$ to obtain
\begin{equation}\label{eq:bathtracedder}
 n(t)  \\ = n(0)+\int_0^t dt' \int_0^t dt''\exv{\timo  \exv{  \mathcal{A}^{L \dagger} (t)\mathcal{L}_I(t')}_B\exv{  \mathcal{A}^{L} (t)\mathcal{L}_I(t'')}_B \text{e}^{\Phi (t)}}_S. 
\end{equation}
To evaluate Eq.~\eqref{eq:bathtracedder} we now use
\begin{equation}\label{eq:genliou}
    \mathcal{L}_I(t')=-ig\mathcal{S}^L(t')\left(\mathcal{A}^{L \dagger}(t')+\mathcal{A}^{ L}(t')\right)+ig\mathcal{S}^R(t')\left(\mathcal{A}^{R \dagger}(t')+\mathcal{A}^{ R}(t')\right),
\end{equation}
 which results in
\begin{align}\label{eq:muchmaths}
n(t)=n(0)+g^2\int_0^t dt' \int_0^t dt'' &\Big[ \exv{\timo \mathcal{S}^L (t') \mathcal{S}^R(t'')}_H\exv{\mathcal{A}^{L \dagger}(t) \mathcal{A}^L(t')}_B \exv{\mathcal{A}^{L}(t)  \mathcal{A}^{R \dagger}(t'')}_B \nonumber \\
&-\exv{\timo \mathcal{S}^L (t') \mathcal{S}^L(t'')}_H \exv{\mathcal{A}^{L \dagger}(t)  \mathcal{A}^L(t')}_B \exv{\mathcal{A}^{ L}(t)  \mathcal{A}^{L \dagger}(t'')}_B \nonumber \\
&+\exv{\timo \mathcal{S}^R (t') \mathcal{S}^L(t'')}_H \exv{\mathcal{A}^{L \dagger}(t)  \mathcal{A}^R(t') }_B \exv{\mathcal{A}^{ L}(t)  \mathcal{A}^{L \dagger}(t'')}_B   \nonumber \\
&- \exv{\timo \mathcal{S}^R (t') \mathcal{S}^R(t'')}_H \exv{\mathcal{A}^{L \dagger}(t)  \mathcal{A}^R(t') }_B \exv{\mathcal{A}^{L}(t)  \mathcal{A}^{R \dagger}(t'')}_B \Big],
\end{align}
where we have used the fact that
\begin{equation}
    \exv{\mathcal{A}^{\alpha \dagger}(t)  \mathcal{A}^{\alpha' \dagger}(t')}_B=\exv{\mathcal{A}^{ \alpha}(t)  \mathcal{A}^{ \alpha'}(t')}_B=0,
\end{equation}
for any $\alpha$ and $\alpha'$. We have also defined
\begin{equation}
    \exv{ \timo \mathcal{S}^\alpha (t') \mathcal{S}^{\alpha'}(t'')}_H=\exv{\timo \mathcal{S}^\alpha (t') \mathcal{S}^{\alpha'}(t'') \text{e}^{\Phi(t)}}_S=\exv{\timo \mathcal{S}^\alpha (t') \mathcal{S}^{\alpha'}(t'') \text{e}^{\int_0^t\mathcal{L}_I(t')dt'}}
\end{equation}
to indicate expectation values taken with respect to the full system density matrix under full system evolution. To be clear, the subscript $H$ means superoperators within the brackets are in the full system Heisenberg picture and the trace is taken with respect to the full system density matrix, the subscript $S$ means superoperators are in the interaction picture and the trace is taken with respect to the reduced system density matrix, and the brackets without subscript mean superoperators are in the interaction picture and the trace is taken with respect to the full system density matrix. For instance, if $t'>t''$ then
\begin{equation}
  \exv{ \timo \mathcal{S}^L (t') \mathcal{S}^{L}(t'')}_H=\exv{ \mathcal{S}^L (t') \mathcal{S}^{L}(t'')}_H 
  =\mathrm{Tr}[ s(t') s(t'')\rho].
\end{equation}
Now we evaluate the bath expectation values in Eq.~\eqref{eq:muchmaths}. For example,
in the first line of Eq.~\eqref{eq:muchmaths} these evaluate to
\begin{equation}
    \exv{\mathcal{A}^{L \dagger}(t) \mathcal{A}^L(t')}_B \exv{\mathcal{A}^{L}(t)  \mathcal{A}^{R \dagger}(t'')}_B=n(0)^2\text{e}^{-i \omega (t'-t'')},
\end{equation}
where we have used cyclicity of the trace, $\exv{\mathcal{A}^{L}(t)  \mathcal{A}^{R \dagger}(t'')}_B=\exv{\mathcal{A}^{L \dagger}(t'')\mathcal{A}^{L}(t)  }_B$, and made the interaction picture time evolution of bath superoperators explicit, $\mathcal{A}^{\alpha}(t)=\mathcal{A}^{\alpha}\text{e}^{-i \omega t}$. The overall result is
\begin{align}\label{eq:waymuchmaths}
n(t)=n(0)+g^2\int_0^t dt' \int_0^t dt'' &\text{e}^{-i \omega (t'-t'')}\Big[ \exv{\timo \mathcal{S}^L (t') \mathcal{S}^R(t'')}_H n(0)^2 \nonumber \\
&-\exv{\timo \mathcal{S}^L (t') \mathcal{S}^L(t'')}_H n(0)(n(0)+1) \nonumber \\
&+\exv{\timo \mathcal{S}^R (t') \mathcal{S}^L(t'')}_H (n(0)+1)^2  \nonumber \\
&- \exv{\timo \mathcal{S}^R (t') \mathcal{S}^R(t'')}_H n(0)(n(0)+1) \Big].
\end{align}
To help us enforce time-ordering upon system superoperators we split the rectangular integral domain into two triangular domains,
\begin{equation}
   \int_0^t dt' \int_0^t dt''=\int_0^t dt' \int_0^{t'} dt''+\int_0^t dt'' \int_0^{t''} dt',
\end{equation}
where the first domain has $t'>t''$ and the second had $t''>t'$. With this, and again using cyclicity of the trace, e.g.  $\exv{\mathcal{S}^L(t') \mathcal{S}^R(t'')}=\exv{\mathcal{S}^L(t'')\mathcal{S}^L(t') }$, we find that terms propotional to $n(0)^2$ cancel out and we are left with
\begin{align}\label{eq:occsupp}
n(t) &= n(0) + g^2 \int_0^t dt' \int_0^t dt'' \Tra [s(t')s(t'')\rho]\{ [n(0)+1]  \text{e}^{-i \omega (t'-t'')}-n(0)  \text{e}^{i \omega (t'-t'')} \}\\
&= n(0) + g^2 \int_0^t dt' \int_0^t dt'' \Tra [s(t')s(t'')\rho](\cos(\omega (t'-t''))-i \sin(\omega(t'-t''))\coth(\omega /2T)),
\end{align}
as used in the main text.

\section{Two-time correlation functions}
\label{sec:apptwomode}
In this section we show it is possible to extend the above approach to not just calculate equal time correlation functions but also calculate general two-point correlation functions of bath operators. For concreteness we'll consider
\begin{equation}\label{eq:twotimecorr}
   C(t,t_1)=\Tra[a^\dagger(t)a(t_1)\rho]= \exv{\timo \mathcal{A}^{L\dagger}(t)A^{L}(t_1)\mathrm{e}^{\int_0^t\mathcal{L}_I(t')\, dt'}}
\end{equation}
where $t>t_1$. The expression in terms of system correlation functions is derived in a similar fashion to the number operator shown in the previous section. It can also be expressed as derivatives of a generating functional as in Eq.~\eqref{eq:ngen2} but with:
\begin{equation}
    \overline{\mathcal{L}_I} (\Lambda,\Lambda^*,t')= \mathcal{L}_I (t') +\Lambda \mathcal{A}^{L}(t') \delta(t-t')+\Lambda^* \mathcal{A}^{L \dagger}(t')\delta(t_1-t').
\end{equation}
The absence of the term $\propto |\Lambda|^2$ that arose in Eq.~\eqref{eq:enterders} is due to the fact that $[\mathcal{A}^{L \dagger}(t), \mathcal{A}^L (t_1)]=0$ under time ordering. Proceeding with the trace over the bath as before gives
\begin{multline}\label{eq:bathtracedcorr}
 \exv{\timo   \text{e}^{\int_0^t \overline{\mathcal{L}_I} (\Lambda,\Lambda^*,t') dt'}}_B   = \exp\left(\Phi (t) + \Lambda^* \int_0^t dt' \exv{  \mathcal{A}^{L \dagger} (t)\mathcal{L}_I(t')}_B + \Lambda \int_0^{t_1} dt' \exv{  \mathcal{A}^{L} (t_1)\mathcal{L}_I(t')}_B +\right. \\ \left. +  \Lambda \int_{t_1}^t dt' \exv{\mathcal{L}_I(t')  \mathcal{A}^{L} (t_1)}_B + |\Lambda|^2 n(0) e^{i\omega (t-t_1)} \right).   
\end{multline}
Taking derivatives with respect to $\Lambda$ and $\Lambda^*$ and setting $\Lambda=\Lambda^*=0$ then allows us to write
\begin{multline}
    C(t,t_1) = \int_0^t dt' \int_0^{t_1} dt''\exv{\timo  \exv{  \mathcal{A}^{L \dagger} (t)\mathcal{L}_I(t')}_B\exv{  \mathcal{A}^{L} (t_1)\mathcal{L}_I(t'')}_B \text{e}^{\Phi (t)}}_S + \\ +  \int_0^t dt' \int_{t_1}^t dt''\exv{\timo  \exv{  \mathcal{A}^{L \dagger} (t)\mathcal{L}_I(t')}_B\exv{  \mathcal{L}_I(t'')\mathcal{A}^{L} (t_1)}_B \text{e}^{\Phi (t)}}_S. 
\end{multline}
The derivation now proceeds by substituting the same form of $\mathcal{L}_I$ as defined in Eq.~\eqref{eq:genliou} and then time-ordering. The time-ordering is done by again splitting the integrals into parts with a fixed ordering i.e.
\begin{equation}
    \int_0^t dt' \int_0^{t_1} dt'' = \int_{t_1}^t dt' \int_0^{t_1} dt'' + \int_0^{t_1} dt' \int_0^{t'} dt'' + \int_0^{t_1} dt'' \int_0^{t''} dt',
\end{equation}
and
\begin{equation}
    \int_0^t dt' \int_{t_1}^t dt'' =  \int_{t_1}^t dt'' \int_0^{t_1} dt' + \int_{t_1}^t dt' \int_{t_1}^{t'} dt'' + \int_{t_1}^t dt'' \int_{t_1}^{t''} dt'.
\end{equation}
The integrands of these terms can then be separately calculated and, after some tedious algebra, re-combined to yield:
\begin{align}\label{eq:twotimeresult}
    C(t,t_1)\mathrm{e}^{-i \omega (t - t_1)} = &n(0)+ g^2 \int_0^{t}dt'\int_0^{t_1}dt''\left\{ (n(0)+1)\Tra[s(t')s(t'')\rho]-n(0)\Tra[s(t'')s(t')\rho]\right\}\mathrm{e}^{-i\omega(t'-t'')}    \nonumber \\
    &+2i g^2\int_{t_1}^{t}dt'\int_{t_1}^{t'}dt'' n(0)\Im{\Tra[s(t')s(t'')\rho]}\mathrm{e}^{-i\omega(t'-t'')}.
\end{align}

Within this framework we are free to consider any free evolution for the bosons which make up the environment. To make the analysis of results simpler and to allow the system to reach a true steady state  such that the correlation functions end up with only one time argument we choose to add dissipation to these environment modes. To be concrete we consider a master equation of the form:
\begin{equation}
    \frac{d}{dt}\sket{\rho(t)} = [\mathcal{L}_I(t)+\gamma_{\downarrow}\mathcal{D}(a)+\gamma_\uparrow \mathcal{D}(a^\dagger)]\sket{\rho(t)},
\end{equation}
where $\gamma_\downarrow$ and $\gamma_\uparrow$ are the bosonic decay and excitation rates respectively. The Lindblad terms are explicitly given by:
\begin{equation}\label{eq:dissdef}
    \mathcal{D}(x) = \mathcal{X}^L\mathcal{X}^{R^\dagger}-\frac{1}{2} \mathcal{X^\dagger}^{L}\mathcal{X}^{L} - \frac{1}{2} \mathcal{X}^{R}\mathcal{X^\dagger}^{R}.
\end{equation}
To simplify the free bath correlation functions we set the transition rates to $\gamma_\downarrow = \gamma[n(0) + 1]$ and $ \gamma_\uparrow = \gamma n(0)$ such that the initial bath state is stationary with respect to the dissipation. The time dependence of the free correlation functions in this case depend on the time-ordering, for example:
\begin{equation}
\langle \timo \mathcal{A}^{L\dagger}(t) \mathcal{A}^L(t')\rangle_B =
    \begin{cases}
\langle  \mathcal{A}^{L\dagger}(0) \mathcal{A}^L(0)\rangle_B\mathrm{e}^{\left(i\omega-\frac{\gamma}{2}\right)(t-t')} & t \geq t' \\
\langle \mathcal{A}^L(0) \mathcal{A}^{L\dagger}(0) \rangle_B\mathrm{e}^{\left(i\omega+\frac{\gamma}{2}\right)(t-t')} & t < t'
\end{cases}.
\end{equation}
Taking this into account results into an equivalent expression to Eq.~\eqref{eq:twotimeresult} down to the phase factors:
\begin{align}\label{eq:twotimeresultdiss}
    C(t,t_1) = &n(0)\mathrm{e}^{\left(i \omega-\frac{\gamma}{2}\right) (t - t_1)} \nonumber \\ &+ g^2 \mathrm{e}^{i\omega(t-t_1)-\frac{\gamma}{2}(t+t_1)} \int_0^{t}dt'\int_0^{t_1}dt'' \left\{ (n(0)+1)\Tra[s(t')s(t'')\rho]-n(0)\Tra[s(t'')s(t')\rho]\right\} \mathrm{e}^{-i\omega(t'-t'')+\frac{\gamma}{2}(t'+t'')}    \nonumber \\
    &+2i g^2 \mathrm{e}^{\left(i\omega-\frac{\gamma}{2}\right)(t-t_1)}\int_{t_1}^{t}dt'\int_{t_1}^{t'}dt'' n(0)\Im{\Tra[s(t')s(t'')\rho]}\mathrm{e}^{-\left(i\omega-\frac{\gamma}{2}\right)(t'-t'')}.
\end{align}

We now wish to take the stationary limit of this expression, that is take $t_1\to \infty$, such that we can write $\lim_{t_1\to \infty}C(t,t_1)\equiv C(t-t_1)$. The first term is unaffected by this as it already only depends on $t-t_1$. Focusing on only the dissipative part of the second term we can express it as:
\begin{equation}
  \int_0^{t}dt'\int_0^{t_1}dt''\dots\mathrm{e}^{-\frac{\gamma}{2}[(t-t')+(t_1-t'')]}.
\end{equation}
From this we can see that the integrand decays as $t-t'$ and $t_1-t''$ increase. If $t_1$ is large enough (and $t$ by the requirement of $t>t_1$) then the system correlation functions, e.g.~$\mathrm{Tr}[s(t')s(t'')\rho]$, will be stationary for all $t'$ and $t''$ where the integrand is of significant weight. If we denote the stationary system correlation functions as
\begin{equation}
    \chi(\tau) =\lim_{t\to\infty} \mathrm{Tr}[s(t+\tau)s(t)\rho]
\end{equation}
and make the shifts $t'\rightarrow t-t'$ and $t''\rightarrow t_1-t''$ we can express this term as
\begin{align}\label{eq:middleterm}
     g^2  \int_0^{t}dt' \int_0^{t_1}dt''  \left\{ [n(0)+1]\chi[(t-t_1)-(t' -t'' )]-n(0)\chi[(t-t_1)-(t' -t'' )]^*\right\} \mathrm{e}^{i\omega(t' -t'' )-\frac{\gamma}{2}(t' +t'' )}.   
\end{align}
The integrand now decays with increasing $t'$ and $t''$ such that in the stationary limit we can take the integral limits to infinity.
Focusing now on the last term of~\eqref{eq:twotimeresultdiss} we can make the shifts $t'\rightarrow t'-t_1$ and $t''\rightarrow t''-t_1$ to yield
\begin{equation}
    2i g^2\int_{0}^{t-t_1}dt'\int_{0}^{t'}dt'' n(0)\Im{\chi(t'-t'')}\mathrm{e}^{-i\omega(t'-t'')}.
\end{equation}
Putting all of this together we can express the stationary bath correlation function as
\begin{align}\label{eq:stationarytwotime}
    C(\tau) = &n(0)\mathrm{e}^{\left(i \omega-\frac{\gamma}{2}\right) \tau}\nonumber \\&+ g^2  \int_0^{\infty}dt'\int_0^{\infty}dt''\left\{ [n(0)+1]\chi[\tau-(t' -t'' )]-n(0)\chi[\tau-(t' -t'' )]^*\right\} \mathrm{e}^{i\omega(t' -t'' )-\frac{\gamma}{2}(t' +t'' )}   \nonumber \\
    &+2i g^2 \mathrm{e}^{\left(i\omega-\frac{\gamma}{2}\right)\tau}\int_{0}^{\tau}dt'\int_{0}^{t'}dt'' n(0)\Im{\chi(t'-t'')}\mathrm{e}^{-\left(i\omega-\frac{\gamma}{2}\right)(t'-t'')}.
\end{align}

\section{Toy model \label{sec:bench}}

\begin{figure}
\centering\includegraphics[width=0.6\linewidth]{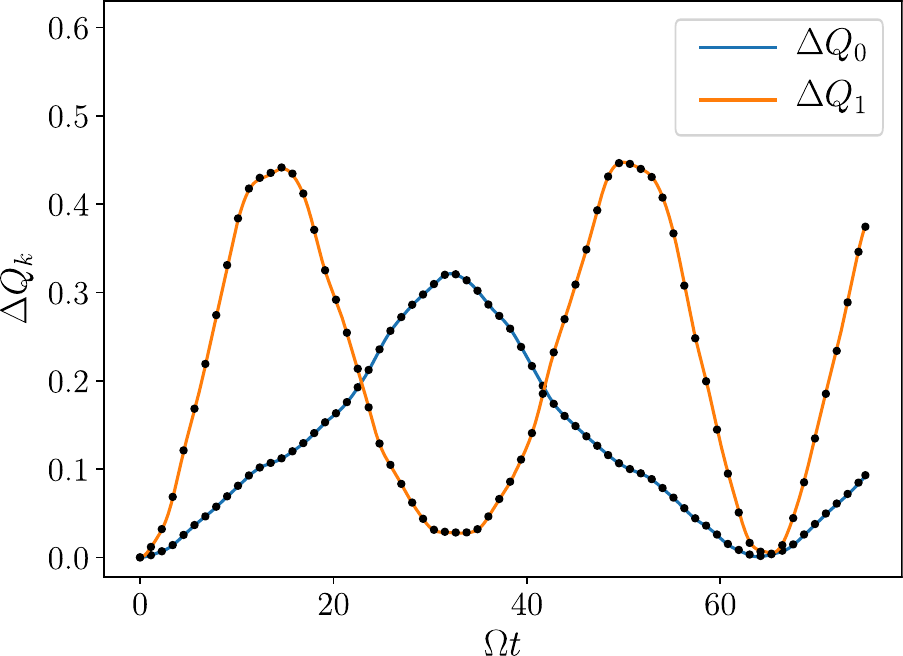}
\caption{\label{fig:benchmark} Comparison of prediction by Eq.~\eqref{eq:occsupp} (dots) with the exact solution (lines) calculated by propagating the full system and environment representing the bosons with a truncated basis of 4 levels each. The other parameters are $\epsilon = 0.1\Omega$, $g_0=0.1\Omega$, $g_1=0.2\Omega$, $\omega_0=0.9\Omega$, $\omega_1=1.1\Omega$ and $T=0.1\Omega$.}
\end{figure}

In this section we benchmark our technique by calculating the energy change in a particular mode, $\Delta Q_q$, for a two-level system coupled to  an environment consisting of only two bosonic modes. With so few degrees of freedom we can simulate the dynamics of the modes explicitly and compare with the prediction given by Eq.~\eqref{eq:occinf}. The model we use is the two-mode Rabi model, the Hamiltonian of which is given by
\begin{align}
H_{R} = \epsilon s_z + \Omega s_x &+ s_z g_0 (a_0 + a_0^\dagger) +  \omega_0 a_0^\dagger a_0 \nonumber \\
&+s_z g_1 (a_1 + a_1^\dagger) +  \omega_1 a_1^\dagger a_1,
\end{align}
and describes a two-level system driven with strength $\Omega$ and bias $\epsilon$. The environment consists of bosons of frequency $\omega_q$ where $a_{q}^{\dagger}$ creates an excitation in the mode indexed by $q\in \{0,1\}$. The system is coupled to each of these modes with strength $g_q$.

In Fig.~\ref{fig:benchmark} we plot the dynamics of each mode, calculated first by numerically integrating the Schr\"odinger equation for the full system-environment dynamics, and then from Eq.~\eqref{eq:occsupp}. We initialised the two-level system in the spin-up state and each bosonic mode in a thermal state at temperature $T=0.1\Omega$. It is clear that our new method correctly predicts the environment dynamics for both modes.

Let us now introduce some dissipation into this model such that the density matrix evolves according to
\begin{equation}
    \frac{d}{dt} \sket{\rho(t)} = \mathcal{L}_R \sket{\rho(t)} + \gamma_\downarrow \mathcal{D}(a)+\gamma_\uparrow \mathcal{D}(a^\dagger),
\end{equation}
where the dissipative terms $\mathcal{D}(\cdot)$ are as defined in Eq.~\eqref{eq:dissdef} and we set $\gamma_\downarrow = \gamma[n(0)+1]$ and $\gamma_\uparrow = \gamma n(0)$.

\begin{figure}
\centering
\includegraphics[width=0.6\linewidth]{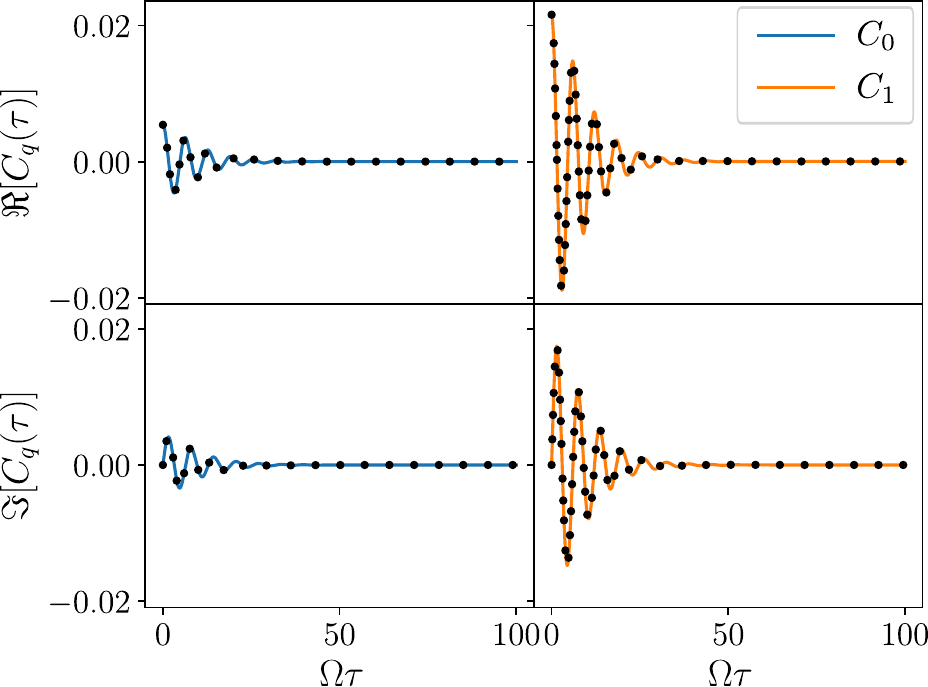}
\caption{\label{fig:benchmark2} Comparison of the prediction of Eq.~\eqref{eq:stationarytwotime} (dots) with the exact solution (lines) calculated by propagating the full system and environment representing the bosons with a truncated basis of 4 levels each. The dissipation rates are set to $\gamma_0 = \gamma_1 = 0.5\Omega$.}
\end{figure}

In Fig.~\ref{fig:benchmark2} we plot the two-time correlation function $C_q(\tau)$ (of each mode $q$) as defined in Eq.~\eqref{eq:stationarytwotime}. The same initial state and temperature were used as the unitary non-dissipative case. We can again see good agreement between our result and the exact dynamics. The correlation functions calculated here can readily be Fourier transformed to calculate emission spectra for these degrees of freedom therefore opening the possibility of calculating an experimentally observable quantity.

\section{Higher order correlation functions}
\label{app:g2}

In calculating higher order correlation functions it is beneficial to express the generating functional in a  more concise way. This can be done in the following way:
\begin{equation}
    G(\mathbf{\Lambda})= \exv{\timo\exp\left[\Phi (t) +\int_0^t \int_0^t  \mathbf{\Lambda}^\dagger(t') \mathbf{X}(t',t'')+\mathbf{\Lambda}^\dagger(t') \mathbf{M}(t',t'') \mathbf{\Lambda}(t'') +  \phi(\mathbf{\Lambda}^\dagger(t'),\mathbf{\Lambda}(t''),t)dt' dt''  \right]}.
\end{equation}
Here we have defined the vectors
\begin{equation}
    \mathbf{\Lambda}(t') = \begin{pmatrix}
    \Lambda^L (t') \\
    \Lambda^R (t') \\
    \Lambda^{L*} (t') \\
    \Lambda^{R*} (t') 
    \end{pmatrix}, \hspace{1cm} \mathbf{X}(t',t'') = \begin{pmatrix}
    \exv{\timo \mathcal{A}^{L\dagger}(t') \mathcal{L}_I (t'')}_B \\
    \exv{\timo \mathcal{A}^{R\dagger}(t') \mathcal{L}_I (t'')}_B \\
    \exv{\timo \mathcal{A}^{L}(t') \mathcal{L}_I (t'')}_B \\
    \exv{\timo \mathcal{A}^{R}(t') \mathcal{L}_I (t'')}_B.
    \end{pmatrix},
\end{equation}
the matrix
\begin{equation}
    \mathbf{M}(t',t'') = \begin{pmatrix}
    \exv{\timo \mathcal{A}^{L\dagger}(t') \mathcal{A}^{L}(t'')}_B & \exv{\mathcal{A}^{R}(t'') \mathcal{A}^{L\dagger}(t')}_B & 0 & 0 \\
    \exv{ \mathcal{A}^{R\dagger}(t') \mathcal{A}^{L}(t'')}_B & \exv{\timo \mathcal{A}^{R\dagger}(t') \mathcal{A}^{R}(t'')}_B & 0 & 0 \\
    0 & 0 & \exv{\timo \mathcal{A}^{L}(t') \mathcal{A}^{L\dagger}(t'')}_B & \exv{\mathcal{A}^{R\dagger}(t'')\mathcal{A}^{L}(t') }_B \\
    0 & 0 & \exv{ \mathcal{A}^{R}(t') \mathcal{A}^{L\dagger}(t'')}_B & \exv{\timo \mathcal{A}^{R}(t') \mathcal{A}^{R\dagger}(t'')}_B
    \end{pmatrix},
\end{equation}
and $\phi$ is the same scalar function as previously which depends on the order of the ladder operators in the correlation function.

It can often be easier to simplify things in terms of these matrix elements and then substitute the explicit values. For example, let's consider the second order coherence at propagation time $t$ with zero time delay, $g^{(2)}(t,0)$ defined as
\begin{equation}
    g^{(2)}(t,0) = \frac{\Tra{\left[a^\dagger(t)a^\dagger(t)a(t)a(t)\rho\right]}}{\Tra{\left[a^\dagger(t) a(t)\rho\right]}^2} =\frac{\exv{\mathcal{A}^{R\dagger}(t) \mathcal{A}^{R\dagger}(t)\mathcal{A}^{L}(t)\mathcal{A}^{L}(t)}}{\exv{\mathcal{A}^{R\dagger}(t)\mathcal{A}^{L}(t)}^2}.
\end{equation}
This form has the advantage of all the superoperators in the numerator and denominator commuting amongst themselves meaning that $\phi = 0$ in both cases. Focusing on the numerator we can express this in terms of the matrix elements as
\begin{align}
    \exv{\mathcal{A}^{R\dagger}(t)\mathcal{A}^{R\dagger}(t)\mathcal{A}^{L}(t) \mathcal{A}^{L}(t)}&= \frac{\delta}{\delta\Lambda^{R*}(t)\delta \Lambda^{R*}(t)\delta \Lambda^{L}(t)\delta \Lambda^L(t)}G(\mathbf{\Lambda})\nonumber \\ &= \int_0^t\left[ \exv{\timo X_1^2 X_2^2} + 4 (M_{10}+M_{23})\exv{\timo X_1 X_2} \right]dt' + 2(M_{10}+M_{23})^2.
\end{align}
Here we have used a simplified notation such that $M_{ij}\equiv M_{ij}(t,t)$ and $X_i\equiv X_i(t,t')$ where $t'$ is a dummy variable integrated from $0$ to $t$ (independently for all $X_i$). In this notation the denominator of $g^{(2)} (t,0)$ can be expressed as
\begin{align}
    \exv{\mathcal{A}^{R\dagger}(t)\mathcal{A}^{L}(t)}^2 &= \left[\frac{\delta}{\delta\Lambda^{R*}(t)\delta \Lambda^{L}(t)}G(\mathbf{\Lambda})\right]^2 \nonumber \\
    &=  \int_0^t \left[\exv{\timo X_1 X_2}^2 + 2 (M_{10}+M_{23})\exv{\timo X_1 X_2}\right] dt' + (M_{10}+M_{23})^2.
\end{align}
By comparing these expressions we can see that $g^{(2)}(t,0)$ is given by
\begin{equation}
    g^{(2)}(t,0) = 2+\frac{\int_0^t \exv{\timo X_1^2 X_2^2}-2\exv{\timo X_1 X_2}^2 }{ \exv{\mathcal{A}^{R\dagger}(t)\mathcal{A}^{L}(t)}^2}.
\end{equation}
For a thermal $\rho_B(t)$ we expect $g^{(2)}(t,0)=2$ and therefore we can identify the second term as a measure of the departure of the bath from a thermal state over time. 
\end{document}